\documentclass[aip,pop,amsmath,amssymb,preprint,tightenlines]{revtex4-1}

\usepackage[pdftex,dvipsnames]{xcolor}

\usepackage[final]{graphicx}
\PassOptionsToPackage{hyphens}{url}
\usepackage[hypertexnames=false]{hyperref}

\usepackage[utf8]{inputenc}
\usepackage[T1]{fontenc}
\usepackage{mathptmx}
\usepackage{etoolbox}

\setcitestyle{authoryear,round,semicolon}

\makeatletter
\def\@email#1#2{%
 \endgroup
 \patchcmd{\titleblock@produce}
  {\frontmatter@RRAPformat}
  {\frontmatter@RRAPformat{\produce@RRAP{*#1\href{mailto:#2}{#2}}}\frontmatter@RRAPformat}
  {}{}
}%
\makeatother

\begin{document}


\title[Accelerated Electrostatic PIC]{Accelerated Steady-State Electrostatic Particle-in-Cell Simulation of Langmuir Probes}




\author{Gregory R. Werner}
\email[]{Greg.Werner@colorado.edu}

\author{Scott Robertson}
\affiliation{Center for Integrated Plasma Studies, University of Colorado, Boulder, Colorado 80309, USA}

\author{Thomas G. Jenkins}
\affiliation{Tech-X Corporation, 5621 Arapahoe Avenue Suite A, Boulder, Colorado 80303, USA}

\author{Andrew M. Chap}
\affiliation{Tech-X Corporation, 5621 Arapahoe Avenue Suite A, Boulder, Colorado 80303, USA}
\affiliation{Currently at AST \& Science, 5825 University Research Ct. \#2300, College Park, MD 20740, USA}

\author{John R. Cary}
\affiliation{Center for Integrated Plasma Studies, University of Colorado, Boulder, Colorado 80309, USA}
\affiliation{Tech-X Corporation, 5621 Arapahoe Avenue Suite A, Boulder, Colorado 80303, USA}



\begin{abstract}
First-principles particle-in-cell (PIC) simulation is a powerful tool for understanding plasma behavior, but this power often comes at great computational expense.
Artificially reducing the ion/electron mass ratio is a time-honored practice to reduce simulation costs.
Usually, this is a severe approximation.
However, for steady-state collisionless, electrostatic (Vlasov-Poisson) systems, the solution with reduced mass ratio can be scaled to the solution for the real mass ratio, with no approximation.
This ``scaled mass'' method, which works with already-existing PIC codes, can reduce the computation time for a large class of electrostatic PIC simulations by the square root of the mass ratio.
The particle distributions of the resulting steady state must be trivially rescaled to yield the true distributions, but the self-consistent electrostatic field is independent of the mass ratio.
This method is equivalent to `numerical timestepping,' an approach that evolves electron and ion populations with different timesteps.
Numerical timestepping can be viewed as a special case of the speed-limited PIC (SLPIC) method, which is not restricted to steady-state phenomena.
Although the scaled-mass approach is simplest, numerical timestepping and SLPIC more easily generalize to include other effects, such as collisions.
The equivalence of these new approaches is demonstrated by applying them to simulate a cylindrical Langmuir probe in electron-argon plasma, speeding up simulation by two orders of magnitude.
Methods such as SLPIC can therefore play an invaluable role in interpreting probe measurements by including geometric effects, collisions, secondary emission, and non-Maxwellian distributions.

\end{abstract}


\maketitle 

\renewcommand{\baselinestretch}{0.1}\normalsize
\tableofcontents
\renewcommand{\baselinestretch}{1.0}\normalsize

\section{Introduction}

This study demonstrates how a large class of steady-state electrostatic plasma simulations can be sped up by the ratio of typical electron and ion velocities, $v_{\rm th,e}/v_{\rm th,i}$,
with almost any existing electrostatic particle-in-cell (PIC) simulation code, simply by using an artificially-low ion mass.
In many cases, using an artificial ion mass $m_i'=m_i (v_{\rm th,i}/v_{\rm th,e})^2$, where $m_i$ is the true ion mass, will yield $v_{\rm th,i}' = v_{\rm th,e}$; reducing the separation between electron and (artificial) ion velocities in this way allows faster simulation.
For example, if ions and electrons have equal temperatures, then 
a speed-up by $v_{\rm th,e}/v_{\rm th,i}=\sqrt{m_i/m_e}$ (where $m_i/m_e$ is the ion/electron mass ratio) can be achieved with an artificial ion mass equal to the electron mass, $m_i'=m_e$.
The reduced mass ratio technique is widely used to speed up plasma simulation when the separation between electron and ion velocity scales inflates the computational cost.
However, it has not been widely appreciated that---for \emph{steady-state electrostatics}---using $m_i'=m_e$ introduces no approximation, as long as the simulated ion velocity distributions (hence also ion currents) are subsequently rescaled using the real mass ratio.
In a practical simulation, the scaled-mass approach may involve a bit more than setting the ion mass to a lower value; velocities of injected particles must always be properly scaled.
For example, if the solar wind is injected into a simulation, one must remember to scale the ion drift velocity as well as the thermal velocity. It may at first seem wrong for the electrons and ions to drift at different speeds, but the steady state solution, after rescaling, will be independent of the scaled mass.

The ``scaled-mass'' method is equivalent to the ``numerical timestepping'' technique of \citet{Jolivet_Roussel-2002}, which has been used in the SPIS PIC code \citep{Roussel_etal-2008}.  [An essentially similar method was previously used by \citet{Serikov_Nanbu-1997} to speed the motion of neutral atoms in direct current and magnetron discharges \citep{Kolev_Bogaerts-2006,Kolev_Bogaerts-2008,Bogaerts_etal-2009}, but seems not to have been extended to charged particle motion.]
Numerical timestepping advances electrons and ions in time with different timesteps (i.e., using a numerical rather than physical timestep); in a steady state, with constant particle distributions and fields, it is of no consequence that different species use different timesteps.
We will show that numerical timestepping and the scaled mass/velocity approach both follow from the same general scaling of mass, velocity, and time in the steady-state Vlasov-Poisson system (\S\ref{sec:VlasovPoisson}). 
The scaled-mass approach can be used immediately with most PIC codes (with post-processing to rescale ion velocities), and has the added attraction that the simulation is physical---it directly models an electron-positron pair plasma.  Therefore, any numerical instabilities introduced by this approach \citep[cf.][]{Jolivet_Roussel-2002} are in fact numerical instabilities found in pair plasma PIC simulations.
On the other hand, the numerical timestepping technique can be more easily generalized to include effects such as secondary emission \citep{Jolivet_Roussel-2002}, collisions \citep{Serikov_Nanbu-1997}, and possibly magnetic fields and other complications routinely handled with standard PIC techniques.

In fact, numerical timestepping is a special case of the speed-limited PIC (SLPIC) method, which we introduced in \citet{Werner_etal-2018}.
Whereas numerical timestepping uses one timestep for all electrons, and another timestep for all ions,
SLPIC introduces an additional (adjustable) approximation that effectively advances
different particles of the same species with different timesteps (e.g., fast electrons are advanced with smaller timestep than slow electrons).
SLPIC offers some additional speed gain; more important, it offers the potential to simulate slow time-dependent phenomena, because all slow particles (whether ions or electrons) can be advanced with the correct timestep, while fast particles are advanced with an artificial timestep.
The SLPIC approach can also generalize standard PIC techniques for simulating secondary emission, collisions, etc. \citep{Jenkins_etal-2021,Theis_etal-2021}.

Whereas the scaled-mass method can often be used without any changes to existing PIC codes, and numerical timestepping can in principle be implemented easily (though in practice may require many changes throughout an existing PIC code), SLPIC is a little more difficult to implement.
SLPIC requires a nontrivial change to the particle-push, and some care in injecting particles into a simulation.  However, even SLPIC does not involve any major changes to the overall structure of the explicit PIC algorithm.

In this paper, we show analytically the equivalence between standard PIC, scaled-mass PIC, and numerically-timestepped PIC (in the steady-state electrostatic limit), and verify the scaled-mass (hence numerically-timestepped) and SLPIC techniques on the problem of Langmuir probe modeling.
Langmuir probes are widely used to measure plasma density and temperature. However, the determination of density and temperature from raw probe data can be fraught with difficulty due to nonideal plasma effects, such as asymmetric 3D probe geometry, particle trapping, collisions, magnetic field, and secondary emission from the probe surface, as well as the fact that an accurate determination may depend on the self-consistent electric potential around the probe.
While difficult to treat analytically, these effects can all be included in PIC simulation, which can therefore help to interpret probe measurements.
Unfortunately, standard PIC simulation can be very time-consuming---in particular because (typically) ions move much more slowly than electrons.
The scaled-mass and SLPIC approaches yield the same results as standard PIC (see \S\ref{sec:results}) but with much less computation.
This work does not attempt to further Langmuir probe theory or to demonstrate agreement between simulation and experiment---but sets the stage to do that in future work.

In the following section, we will briefly introduce Langmuir probes and describe why Langmuir probe simulations could greatly benefit from faster simulation using scaled-mass and SLPIC techniques;
to be explicit about potential benefits, we quickly review PIC simulation costs in \S\ref{sec:PICcost} so we can state precisely how and when these techniques can save computation time.
The main analytical argument---the mass/velocity/time-scaling of the Vlasov-Poisson system---is presented in \S\ref{sec:VlasovPoisson} to show the equivalence of the scaled-mass and numerical timestepping methods.
Then, in \S\ref{sec:SLPIC} the SLPIC method is briefly reviewed.
Test simulations and results for a cylindrical (2D) Langmuir probe are described in \S\ref{sec:setup} and \S\ref{sec:results}, demonstrating that scaled-mass PIC and SLPIC methods are substantially faster than PIC, but just as accurate.
With faster methods, we are able to explore in~\S\ref{sec:trapping} the surprisingly slow convergence to a steady state, which we attribute to particle trapping; although the simulations quickly reach an approximate steady state after a few ion-crossing times, the potential $\Phi(x,y,t)$ continues to change monotonically by $O(0.05T/e$) over thousands of ion-crossing times for a large probe voltage, $V=20 T/e$.
Finally, we conclude with a summary (\S\ref{sec:conclusion}).

\section{Application: Langmuir probes}
\label{sec:probes}

Briefly and simplistically, a Langmuir probe is a conductor (e.g., a small metal wire, disc, or sphere) extended into a plasma, usually at the end of a thin insulated rod.
A bias voltage $V$ is applied to the probe in such a way that the current $I$, required to maintain $V$, can be measured (the supplied current $I$ must equal the plasma current hitting the probe to maintain a steady state); the bias is stepped or swept through a range of voltages while measuring $I$ to yield $I(V)$.
One typically assumes a probe/plasma model that allows calculation of $I(V)$ from the ambient plasma density $n_0=n_e=n_i$ and temperatures ($T_e$ and $T_i$); ultimately, one must find the plasma parameters $n_0$, $T_e$, and possibly $T_i$ such that the model (i.e., in the most general case, the simulation of) $I(V;n_0,T_e,T_i)$ is as close to the measured $I(V)$ as possible.
For a thorough introduction to Langmuir probes, we refer the reader to some of the many works on the subject 
\citep[e.g.,][]{MottSmith_Langmuir-1926,Tonks_Langmuir-1929,Allen_etal-1957,Bernstein_Rabinowitz-1959,Laframboise-1966,Allen-1992,Demidov_etal-2002,Merlino-2007}.

For the purposes of this work, we will consider a cylindrical Langmuir probe in an infinite 2D unmagnetized electron-ion plasma, in the regime $r_p \lesssim \lambda_D \ll \lambda_{\rm mfp}$, where the probe radius $r_p$ is small compared to the Debye screening length 
$\lambda_D$, which is much smaller than the mean free path $\lambda_{\rm mfp}$ (for electrons or ions).
This is the regime most amenable to basic PIC simulation---a collisionless plasma in which the Debye length must be resolved.
Although PIC simulation---and hence this method---can be applied to other regimes, we leave such demonstration to future work.
Since we are not considering chamber walls, we measure all potentials
with respect to the plasma potential far from the probe (at the simulation boundary); i.e., $V=0$ means the probe bias equals the plasma potential (in practice, of course, one
does not know the plasma potential {\it a priori}).

When the probe is smaller than the Debye length, the self-consistent electric field around the probe affects the current $I$ collected by the probe at a given bias voltage $V$.
The orbital-motion-limited (OML) theory \citep[cf.][]{Laframboise-1966} was developed to calculate $I(V)$ by determining which electrons and ions in the ambient plasma far from the probe would reach the probe (where they would be absorbed), assuming collisionless trajectories in the self-consistent electric field (i.e., the field resulting from the applied bias voltage as well as screening effects from the surrounding plasma).
For symmetric probes, such as a cylindrical (wire) or spherical probe, in collisionless plasma, the conservation of energy and angular momentum greatly simplifies calculation of $I(V)$ \citep[e.g.,][]{MottSmith_Langmuir-1926,Laframboise-1966,Allen-1992}.
Symmetry reduces trajectory calculation to 1D motion in a pseudopotential; a particle with more energy than the (angular-momentum-dependent) pseudopotential at the probe will reach the probe, as long as no barrier intervenes.  Whether angular momentum can raise a barrier above the probe pseudopotential depends on the self-consistent $\Phi(r)$ \citep{Allen-1992}. 
When barriers are absent or negligible \citep{Lampe-2001}, $I(V)$ is conveniently independent of the detailed form of~$\Phi(r)$.
Assuming this to be case,
the electron current for a cylindrical probe in a collisionless Maxwellian plasma is 
\citep[e.g.,][]{Allen-1992}
\begin{eqnarray} \label{eq:IofV}
  I_{e}(V) &=& 
   \left\{
   \begin{array}{c@{\quad}l}
  (-e) A_p \frac{n_0 v_{\rm th,e}}{\sqrt{2\pi}} 
               \exp \left[ -(-e)V/m_e v_{\rm th,e}^2 \right]
  & V\leq 0
  \\
  (-e) A_p \frac{n_0 v_{\rm th,e}}{\sqrt{2\pi}} 
               \left[ \frac{2}{\sqrt{\pi}}
                       \sqrt{-\frac{(-e)V}{m_e v_{\rm th,e}^2}}
        + \exp \left( -\frac{(-e)V}{m_e v_{\rm th,e}^2}  \right)
        \textrm{erfc} \left( \sqrt{-\frac{(-e)V}{m_e v_{\rm th,e}^2}}\,  \right)
           \right]
  & V\geq 0
  \end{array}
  \right.
\end{eqnarray}
where $A_p$ is the probe area and
where $v_{\rm th,e} = \sqrt{T_e/m_e}$ is the electron thermal velocity.
Sometimes an approximation is made to simplify this result:
$2\sqrt{\eta/\pi} + \exp (\eta) \textrm{erfc}\sqrt{\eta} \approx
 (2/\sqrt{\pi})\sqrt{1+\eta}$ for $\eta = eV/m_e v_{\rm th,e}^2 \gtrsim 2$ \citep{Allen-1992}.
Analogous expressions can be found for planar and spherical probes.
We note that $J_{e0}=n_0 v_{\rm th,e}/\sqrt{2\pi}$ is the one-way number current density for electrons in a stationary Maxwellian distribution;
i.e., if the probe were at the plasma potential $V=0$ and magically transparent (with no effect on the plasma at all), then $J_{e0}A_p$ is the current that would hit the outside of the probe surface.

Although the ion current $I_i$ in the collisionless regime can be calculated (at least in principle) in exactly the same way as $I_e$, in practice we often have $T_i \ll T_e$, and the ions can be significantly affected by electric fields in the pre-sheath (in nearly quasineutral plasma) before entering the probe sheath of thickness $\sim \lambda_D$.
For example, for $V\lesssim -T_e/e \ll -T_i/e$, 
the pre-sheath may accelerate the ions to 
some fraction of the Bohm velocity $v_B\equiv \sqrt{T_e/m_i}$ \citep[e.g.,][]{ChenIntroPlasma,Merlino-2007,Robertson-2013}.
When the ions enter the non-neutral sheath, they have a directed velocity, which leads to a different calculation from the above, which is for a stationary Maxwellian \citep{Allen_etal-1957,Bernstein_Rabinowitz-1959,Chen-1965,Laframboise-1966}.
Such calculations generally must solve approximately for the self-consistent potential in the pre-sheath.
We mention this just to illustrate how self-consistent numerical simulation is required for more complicated geometries and plasma regimes; for the purpose of this work, however, we stick to the simplest treatment with $T_i=T_e$ in a collisionless plasma, so that we can apply Eq.~(\ref{eq:IofV}) both to electrons and to ions in order to verify simulation results against analytical theory.

Therefore, for a sufficiently symmetric probe placed into an infinite, stationary, Maxwellian, collisionless plasma with zero magnetic field, assuming the probe absorbs all particles that hit it, without secondary emission, and assuming that the resulting potential $\Phi(r)$ does not fall off (is not screened out) too rapidly from the probe, we can write down $I(V)$ in terms of $n_0$, $v_{\rm th,e}$, and $v_{\rm th,i}$ [e.g., for a cylindrical probe, Eq.~(\ref{eq:IofV})].  The expressions are simple enough that, given $I(V)$ over a sufficient range, we can extract $n_0$, $v_{\rm th,e}$, and $v_{\rm th,i}$ (hence also $T_e$ and $T_i$).

PIC simulation is often used to model effects that violate the conditions required for Eq.~(\ref{eq:IofV}).
Even for an ideal cylindrically- or spherically-symmetric probe in a collisionless Maxwellian plasma, Eq.~(\ref{eq:IofV}) can be inaccurate when (pseudo)potential barriers arise due to angular momentum \citep{Lampe-2001}.
Moreover, many non-ideal conditions defeat analytical and
non-kinetic computational approaches.
For example, collisions can significantly affect ion current to Langmuir probes \citep[e.g.,][]{Schulz_Brown-1955,Zakrzewski_Kopiczynski-1974,Sudit_Woods-1994,Sternovsky_Robertson-2002,Sternovsky_etal-2003},
and a number of studies have used PIC codes with Monte Carlo collision capabilities to study this, using 1D cylindrical or spherical models (with 3D velocities) and a Boltzmann electron distribution
\citep{Taccogna_etal-2004,Tejero-del-Caz_etal-2016,Voloshin_etal-2015}
or fully kinetic electrons
\citep{Soberon-2006,Cenian_etal-2005,Iza_Lee-2006,Trunec_etal-2015,Zikan_etal-2019}, including surface effects such as secondary electron emission \citep{Cenian_etal-2014}.
PIC has been critical in simulating Langmuir probes with asymmetric 3D geometries \citep{Seran_etal-2005,Hilgers_etal-2008,Hruby_Hrach-2010,Chiaretta-2011,Imtiaz_etal-2013,Podolnik_etal-2018},
and has been used for probe and sheath analysis in the presence of magnetic fields \citep{Bergmann-2002,Podolnik_etal-2018}.
The effect of non-uniform plasma and flowing plasma conditions are other examples ripe for exploration with PIC \citep{Seran_etal-2005,Knappmiller_Robertson-2007,Olson_etal-2010,Imtiaz_etal-2013}.
When Langmuir probe measurements are complicated by effects such as described above, $I(V)$ cannot often be precisely predicted analytically; simulations can predict $I(V)$, but solving the inverse problem---finding the $n_0,T_e,T_i$ that yield a simulated $I(V)$ closest to the measured $I(V)$---is extremely expensive, computationally.
The techniques presented in this paper could speed Langmuir probe simulation by two orders of magnitude, which makes the inverse problem much more feasible.
Many of the above-mentioned studies could have benefited significantly from the techniques described in this paper; only a few took advantage of numerical timestepping \citep{Seran_etal-2005,Hilgers_etal-2008,Imtiaz_etal-2013}.

\section{PIC simulation cost}
\label{sec:PICcost}

As we vary the ion mass while leaving other physical and numerical parameters fixed,
the cost of many steady-state PIC simulations with electrons and ions 
scales as $C\sim v_{\rm th,e}/v_{\rm th,i}$, the ratio of ion and electron thermal (or typical)
velocities.  If the plasma is thermal with $T_i=T_e$, then $C\sim \sqrt{m_i/m_e}$.
Even for an electron-proton plasma with $T_i=T_e$, this ratio is 40; 
for a plasma with heavier ions and/or hotter electrons, this ratio can be even larger.
E.g., for an electron-argon plasma with $T_i=T_e$, this ratio is 270.

To be more specific, we write down, in rough but general terms, 
the computational cost $C$ of a
$d$-dimensional simulation with volume $L^d$ resolved by 
grid cells of size $\Delta x$, 
total duration $T$ resolved by timestep $\Delta t$, 
and $N_{\rm ppc}$ particles per cell.
Assuming that $v_{\rm th,i} \leq v_{\rm th,e}$, it is often more useful to express $C$ in terms of 
\begin{itemize}
  \item $N_c \equiv v_{\rm th,i} T/L$,
     the number of ion-crossing times required to reach a steady state
     (typically $N_c > 1$);
  \item $f_{\Delta t} \equiv v_{\rm th,e} \Delta t / \Delta x$, the 
    fraction of a grid cell crossed by a typical electron in one timestep
    (typically, $f_{\Delta t} \lesssim 1/3$ to prevent most electrons from traveling more than $\Delta x$ in one timestep);
    and
  \item $\lambda_{De}$, the electron Debye length (typically, $\Delta x \lesssim \lambda_{De}$ is required to avoid unphysical finite-grid heating).
\end{itemize}
For many if not most PIC simulations, 
the cost is proportional to the number of 
particles times the number of timesteps:
\begin{eqnarray}
  C & \propto & N_{\rm ppc} 
    \left( \frac{L}{\Delta x} \right)^d \frac{T}{ \Delta t  }
  =
  N_{\rm ppc} \frac{N_c}{f_{\Delta t}}  
      \left(
      \frac{L}{\lambda_{De}} \right)^{d+1}
      \left(
       \frac{\lambda_{De}}{\Delta x} \right)^{d+1}
       \frac{v_{\rm th,e}}{v_{\rm th,i}}
\end{eqnarray}
which shows that $C \propto v_{\rm th,e}/v_{\rm th,i}$.

We note that, because the electron plasma 
frequency $\omega_{pe}$ is
$\omega_{pe} = v_{\rm th,e}/ \lambda_{De}$, 
the condition $f_{\Delta t} < 1$ ensures that the plasma oscillation
period is resolved ($\omega_{pe}\Delta t < 2$), as long as the Debye
length is resolved better than $\Delta x < 2\lambda_D$.
(Typically, explicit PIC simulations need to resolve plasma oscillations and the Debye length to be stable.)

In this paper, we focus on eliminating the 
considerable factor of $v_{\rm th,e}/v_{\rm th,i}$ in the simulation cost.
However, writing out other terms in the cost helps to show where
the described techniques might be most useful.
For example, if the simulation domain is much larger than the Debye length and there is reason to believe that Debye-length phenomena are unimportant, then other methods may be more appropriate.

\section{Steady state Vlasov-Poisson scaling}
\label{sec:VlasovPoisson}

A solution of the Vlasov-Poisson equations
for particles of mass $m$ can be trivially scaled to find the solution of a different set of Vlasov-Poisson equations for mass $m'$.
The Vlasov-Poisson system for one species, with charge $q$, mass $m$, velocity distribution $f(\mathbf{x},\mathbf{v},t)$, and electric potential $\phi(\mathbf{x},t)$ is:
\begin{eqnarray}
  \partial_t f(\mathbf{x},\mathbf{v},t)
 &=&
  -\mathbf{v} \cdot \partial_{\bf x} f(\mathbf{x},\mathbf{v},t)
  +(q/m)\partial_{\bf x} \phi(\mathbf{x},t)  \cdot 
        \partial_{\bf v} f(\mathbf{x},\mathbf{v},t)
        \\
   -\partial_{\bf x}\cdot \partial_{\bf x} \phi 
   &=& \frac{qn}{\epsilon_0}
   = \frac{q}{\epsilon_0} \int f(\mathbf{x},\mathbf{v},t) d^3 v
\end{eqnarray}
where $\partial_{\bf x}$ and $\partial_{\bf v}$ are gradient operators
in position and velocity space.

Suppose a distribution $f(\mathbf{x},\mathbf{v},t)$ and potential $\phi(\mathbf{x},t)$ satisfy the Vlasov-Poisson system for a species of mass $m$.
If, given any mass $m'$, we scale velocity and time according to
$\mathbf{v}'=\sqrt{m/m'}\,\mathbf{v}$ and $t'=\sqrt{m'/m}\: t$,
then straightforward substitution shows that the scaled
distribution
\begin{eqnarray}
  f'(\mathbf{x},\mathbf{v}',t') &\equiv & 
  \left(\frac{m'}{m} \right)^{3/2}
  f\left(\mathbf{x},
         \mathbf{v}=\sqrt{\frac{m'}{m}} \, \mathbf{v}', 
         t=\sqrt{\frac{m}{m'} }\, t' \right)
\end{eqnarray}
satisfies the scaled Vlasov equation for a species of mass $m'$:
\begin{eqnarray}
  \partial_{t'} f'(\mathbf{x},\mathbf{v}',t')
 &=&
  -\mathbf{v}' \cdot \partial_{\bf x} f'(\mathbf{x},\mathbf{v}',t')
  +(q/m')\partial_{\bf x} \phi  \cdot 
        \partial_{{\bf v}'} f'(\mathbf{x},\mathbf{v}',t')
\end{eqnarray}
Furthermore, the scaled number density (hence also charge density) is unchanged
\begin{eqnarray}
  n'(\mathbf{x},t') &\equiv & \int f'(\mathbf{x},\mathbf{v}',t') d^3 v'
   = \int f(\mathbf{x},\mathbf{v},t) d^3 v = n(\mathbf{x},t)
\end{eqnarray}
so that the scaled solution self-consistently satisfies the
Poisson equation with the same potential $\phi'(\mathbf{x},t')=\phi(\mathbf{x},t)$.
The current density, however, does change; it scales like velocity:
\begin{eqnarray}
  \mathbf{J}'(\mathbf{x},t') &\equiv & 
   \int \mathbf{v}' f'(\mathbf{x},\mathbf{v}',t') d^3 v'
   = \sqrt{\frac{m}{m'}} 
     \int \mathbf{v} f(\mathbf{x},\mathbf{v},t) d^3 v = 
   \sqrt{\frac{m}{m'}} \, \mathbf{J}(\mathbf{x},t)
.\end{eqnarray}

The Vlasov-Poisson system scales similarly for multiple species---except
that to preserve self-consistent time evolution, both species must scale
time in the same way, which does nothing to reduce the scale separation between species.
However, 
in the steady state limit, $\partial_t f, \partial_{t'} f' \rightarrow 0$,
time evolution ceases, and it no longer matters that different species evolve in time at different rates.
Therefore, we can evolve different species, each with an arbitrary mass/velocity/time scaling, and if the solution reaches a steady state, then the result is exactly equivalent to the unscaled system.

Therefore, if we wish to simulate a Vlasov-Poisson system with electrons and ions, we can instead simulate a Vlasov-Poisson system with ions with mass reduced by $m_i'/m_i$ while simultaneously scaling ion velocities by $\sqrt{m_i/m_i'}$.  If we choose $m_i'=m_e$, this will be equivalent to
an electron-positron simulation.
The time evolution of the ions will be too fast by $\sqrt{m_i/m_i'}$ (i.e., positrons move faster than ions), but once we reach the steady state, that no longer matters.
Assuming we reach a steady state, we can scale the ion velocities to their true values by multiplying by $\sqrt{m_i'/m_i}$, and correspondingly any ion currents measured must also be scaled by $\sqrt{m_i'/m_i}$.
However, the potential $\phi(\mathbf{x})$ and the number/charge densities
are correct without any scaling.

We can think of this approach as simulating positrons instead of ions (with velocities scaled appropriately), or equivalently as scaling time for ions, such that ions experience time faster than electrons.
I.e., a positron follows exactly the same spatial trajectory as an ion that
experiences ``fast time'' (sped up by a factor $\sqrt{m_i/m_e}$) and at every point, the positron velocity equals the ion velocity times $\sqrt{m_i/m_e}$.
The latter approach is numerical timestepping \citep{Serikov_Nanbu-1997,Jolivet_Roussel-2002}:
in each simulation step, the electrons evolve by some time $\Delta t$ and the ions by a longer time $\sqrt{m_i/m_i'}\,\Delta t$.
It is important to note that there is no guarantee that the system will reach a steady state, and it is conceivable that the standard PIC simulation might reach a a steady state while the scaled or numerically-timestepped simulation does not, due to some instability \citep{Jolivet_Roussel-2002}.
However, the scaled-mass approach shows that the simulated time evolution is physical (in the sense that it is a valid time-dependent PIC simulation of an electron-positron plasma; it is not a valid time evolution of the system with real ion mass $m_i$).   Therefore, any instability can be understood  as an instability in a classic PIC simulation of electron-positron plasma.

Simulating a Langmuir probe in electron-argon plasma with $T_e=T_i$ should be $\sqrt{m_i/m_e} \approx 270$ times faster than a standard PIC simulation with the real ion mass, and will yield equivalent results.

At first glance, however, the results may not always seem to be equivalent.
For example, the floating potential of a metal object in an (equal-temperature) electron-positron plasma is $V=0$ (the plasma potential),
because electrons and positrons are equally mobile, but the floating potential in an electron-ion plasma is nonzero because electrons are more mobile than ions.
If the potential $\phi(\mathbf{x})$ is independent of the mass as we have shown, how can we have a disagreement about the value of the floating potential?
This is easily resolved: the floating potential is defined as the potential at which the net current to the metal object is zero.
Simulation will rightly show that in electron-positron plasma, the floating potential is 0~V, meaning that at this voltage the electron and positron currents cancel.
When we scale this to find the solution for an electron-ion plasma, the potential does not change; however, the ion current does change (by $\sqrt{m_i'/m_i}$), so that it no longer cancels the electron current.
Therefore, we conclude from the scaled simulation that 0~V is not the floating potential in the electron-ion plasma, which is correct.
If we wanted to find the floating potential in the electron-ion plasma,
we would have to find the $V$ at which the positron (number) current
is $\sqrt{m_i/m_i'}$ times greater than the electron (number) current.

We note that the scaled-mass approach does not apply straightforwardly in the presence of a magnetic field: the magnetic force (unlike the electric force) is proportional to $\mathbf{v}$, and that ruins the scaling.
(Intuitively, one may recognize that the Larmor radius of a particle in
the magnetic field---hence the particle's path---depends on the charge-to-mass ratio.)
This can be rectified if the magnetic field is scaled so that
$\mathbf{v'}\times \mathbf{B}' = \mathbf{v}\times \mathbf{B}$ (this requires scaling $\mathbf{B}$ by the same factor as $\mathbf{J}$), but then we run into the problem that electrons and ions share the same magnetic field, so we cannot rescale ions alone.
Of course, we might be able to modify a PIC code so that electrons and ions would see magnetic fields different by a constant factor
(however, this simulation might be unphysical, with electrons and ions seeing different magnetic fields; in contrast, the rescaled Vlasov-Poisson steady-state system is physical).

On the other hand,
numerical timestepping should work with magnetic fields, because it does not require rescaling velocities.
In general, it is easier to see how one might apply numerical timestepping to  effects such as magnetic fields, collisions, and secondary emission.
Such considerations led us to develop the speed-limited PIC method.

\section{SLPIC}
\label{sec:SLPIC}

In \citet{Werner_etal-2018}, we introduced speed-limited PIC (SLPIC) simulation to manage large differences in velocity scales, e.g., between
electrons and ions, but also within a single species.
This method introduces a slowing-down factor $\beta$ for each particle
[generally $\beta=\beta(v)$ depends on the particle's speed], such that
the equations of motion are changed to
\begin{eqnarray}
  \dot{\mathbf{x}} &=& \beta(v) \mathbf{v} \\
  \dot{\mathbf{v}} &=& \beta(v) \mathbf{a}
\end{eqnarray}
where $\mathbf{a}$ is the acceleration, e.g., $-(q/m)\partial_{\bf x} \phi$
in an electrostatic simulation.
One can see that $\beta$ essentially modifies the time derivative, slowing 
down motion if $\beta < 1$, or speeding it up if $\beta > 1$.

If $\beta$ is independent of velocity, then it uniformly scales time, and the result is equivalent to numerical timestepping \citep{Serikov_Nanbu-1997,Jolivet_Roussel-2002}, or the scaled mass and velocity approach previously described.
However, when $\beta$ depends on velocity, it can reduce scale separations between typical and maximum velocities, including between ion and electron velocities.
In addition, $\beta(v)$ can be chosen so that $\beta=1$ for sufficiently slow particles; this allows slow time-dependent behavior to be simulated \citep{Werner_etal-2018,Jenkins_etal-2021}.  If $\beta=1$ for all particles, then SLPIC is the same as PIC.

It was shown by \citet{Werner_etal-2018} that SLPIC simulation, with velocity-dependent $\beta(v)$ that approaches 1 for small $v$,
essentially solves an approximate 
Vlasov equation.  The approximation becomes increasingly accurate
as fields and particle distribution functions change more slowly in time---and indeed, the approximation
becomes exact in the steady-state limit.
\citet{Jenkins_etal-2021} derived the dispersion relation for Langmuir and ion-acoustic modes in a 1D speed-limited plasma with velocity-dependent $\beta$, and found sufficiently slow wave behavior to be accurately rendered.
When $\beta$ is constant, however, the SLPIC method is equivalent to numerical timestepping and is accurate only in the steady-state limit---even very slow time dependence may be inaccurately simulated.

\section{Simulation setup}
\label{sec:setup}

\begin{figure}
\centering
\includegraphics*[width=4in]{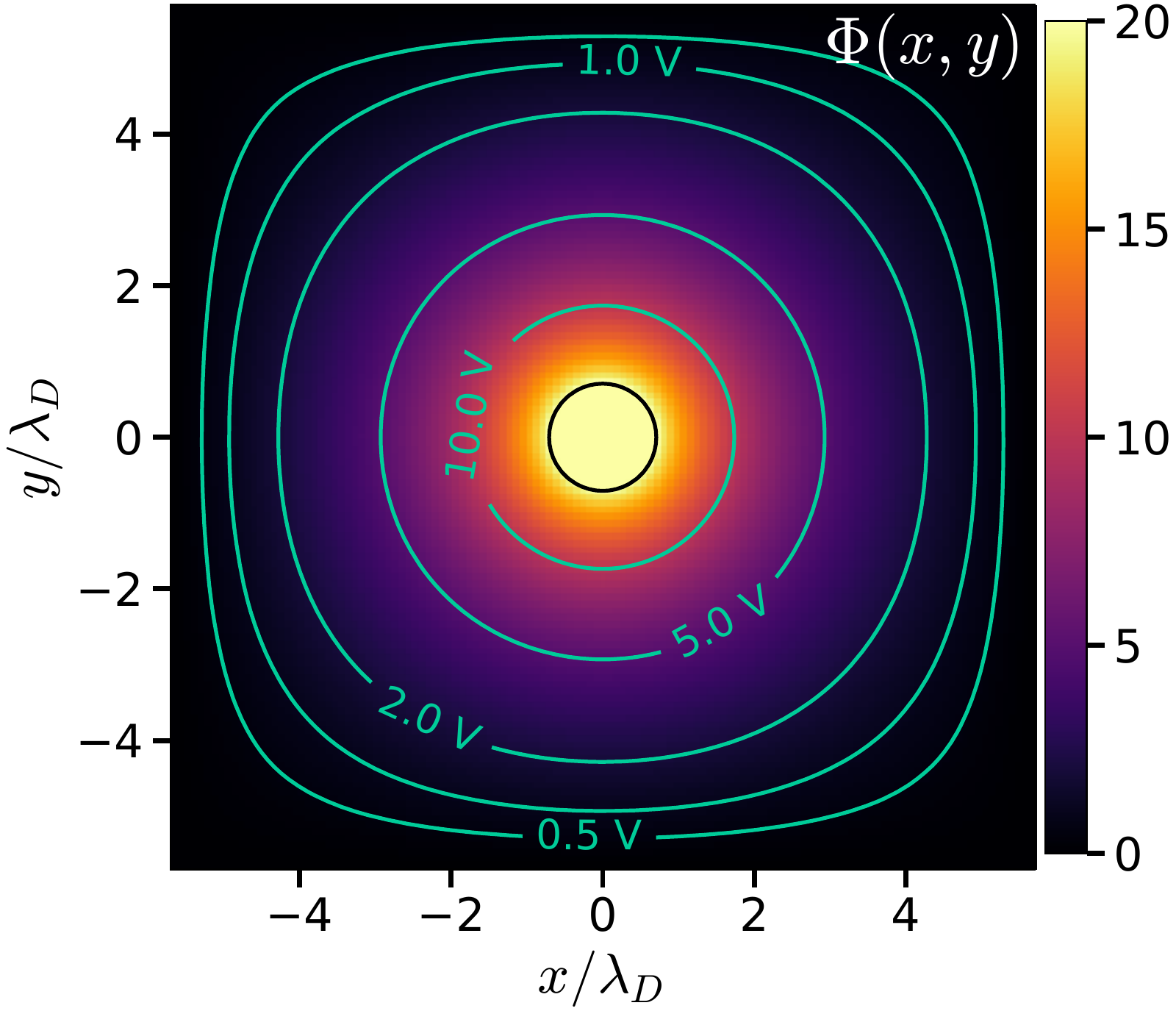}%
\caption{\label{fig:phi2d}
The electric potential $\Phi(x,y)$ (in Volts) for probe voltage~$V=20\:$V, in the electron-positron PIC simulation.
}
\end{figure}

To demonstrate the methods presented in this paper, we used the {\sc Vorpal/VSim} computational framework \citep{Nieter_Cary-2004} to simulate a cylindrical Langmuir probe in 2D using explicit electrostatic PIC simulation with zero magnetic field.
The probe of radius $r_p=0.5\:$cm is centered in a square box of size 
$L=8\:$cm, gridded with $128\times 128$ cells, hence $\Delta x = (1/16)\:$cm.
Dirichlet (conducting) boundary conditions
are applied with $\Phi=0$ at the simulation boundary, and $\Phi=V$
at the probe boundary; the square box breaks perfect cylindrical symmetry, but the box is large enough that this has negligible effect (see~Fig.~\ref{fig:phi2d}).
Electrons and ions (or positrons) are injected from the simulation boundary with stationary Maxwellian distributions of temperature $T=T_i=T_e=1\:$eV and density $n_e=n_i=n_0$ so that the Debye length is $\lambda_{D}=\sqrt{T_e/4\pi (2n_0) e^2} = (1/\sqrt{2})\:$cm.
Particles hitting the simulation or probe boundary are absorbed.
The probe current is measured by counting the (total charge of) 
macroparticles hitting the probe---and, in the case of scaled-mass PIC, scaling the collected positron current by $\sqrt{m_e/m_i}$.
Relevant dimensionless lengths are: $r_p/\lambda_{D} = 0.71$,
$L/\lambda_{De}=11$, $\Delta x/\lambda_{D}=0.088$.

Each species is injected in two sub-populations: one component, forming the bulk of the Mawellian, has particles of equal weight, while the other component, forming the high-energy tail, has particles of varying weights (the weight is the number of physical particles represented by a simulated macroparticle).
The variable-weight particles allow better representation of the high-energy tail, which is necessary to get a sufficient number of particles hitting the probe to calculate the current with statistical significance.
E.g., for $V=20 T/e$, (positively-charged) particles need kinetic energy $E=20 T$ to get near the probe; such particles are rare---one in a billion.
Since a simulation, e.g., with $L=8\:$cm and $\Delta x=(1/16)\:$cm, might have around $5\times 10^5$ macro-ions---around 30 macro-ions per cell on average---we might expect just one ion with $E>20T$ to appear in the simulation every thousand ion-crossing-times.
Particle injection is described in much more detail in Appendix~\ref{sec:injection}.

For standard PIC simulations, particle injection resulted in about 37 particles per cell for a probe voltage $V=0$ (i.e., nearly zero electric field in the simulation).  For nonzero $V$, the injection of the probe-attracted species was the same, resulting in fewer particles per cell because particles speed up closer to the probe, as low as 26 per cell for $|V|=20 T/e$, averaged over all cells.
The injection of the probe-repelled species resulted in 26--32 particles per cell, except for 16 particles per cell for $|V|=20 T/e$ (again, averaged over all cells); the fraction of variable-weight particles ranged from 0.29 to 0.55.
PIC simulations with $m_i=m_e$ had very similar numbers of particles per cell.
In SLPIC simulations, the number of electrons for $V\geq 0$ increased with $V$ from 37--48; for $V<0$, 34--47.
The number of ions for $V\leq 0$ ranged from 45--49; for $V>0$, 38--55.

Measuring probe currents at 20$\:$V requires simulating particles
with 20$\:$eV of energy; such particles have 
$v = \sqrt{40} v_{\rm th} \approx 6.3 v_{\rm th}$, 
where $v_{\rm th}=\sqrt{T/m}$ is the thermal velocity.
The number of particles (from a 1$\:$eV population) 
with higher energies decreases extremely rapidly with energy, so we 
simulated speeds only up to about $7 v_{\rm th}$.

Although a real probe might sweep $V$ through small, equally-spaced steps, for demonstration purposes we simulated $V\in \{-$20, $-$15,$-$10, $-$5, $-$2, $-$1, $-$0.5, 0, 0.5, 1, 2, 5, 10, 20$\}\:$Volts, for a time $T=8L/v_{\rm th,i}$ at each probe voltage to reach a steady state.
The analysis of collected current was performed only in the last
half (time $T/2$) of each simulation to capture the steady-state result. 
Simulating for longer times did not significantly affect the measured probe current (but see~\S\ref{sec:trapping}).

For standard PIC and scaled-mass PIC simulations, 
the requirement that the fastest electron cross less than one cell per timestep led us to use a timestep
$\Delta t = 0.9 \Delta x / (7 v_{\rm th,e})$.
For SLPIC, we chose the speed-limiting function 
$\beta(v) = 1$ for $v\leq 2v_{th,i}$ and
$\beta(v)=2 v_{th,i}/v$ for $v\geq 2v_{th,i}$. 
For this choice of $\beta(v)$, all particles with speeds below $2v_{th,i}$ are simulated exactly as in normal PIC; thus most ions are simulated normally.
However, faster particles are speed-limited so that they 
never travel more than a distance $2v_{th,i}\Delta t$ in any single timestep \citep{Werner_etal-2018}.
This allowed the SLPIC simulation to use 
$\Delta t = 0.9 \Delta x / (2 v_{\rm th,i})$.

For the simulated time $T = 8 L/v_{\rm th,i}$,
the standard electron-argon PIC simulation required
$N_t = (8/0.9) (L/\Delta x) 7v_{\rm th,e}/v_{\rm th,i} \approx 
2\times 10^6$ timesteps.
The electron-positron PIC simulation, with $v_{\rm th,i}=v_{\rm th,e}$,
required only $8\times 10^3$ timesteps---shorter by a factor 
$(40\times 1836)^{1/2} = 270$.
And SLPIC required only $2\times 10^3$ timesteps.

\section{Simulation results}
\label{sec:results}

\begin{figure}
\centering
\includegraphics*[width=3in]{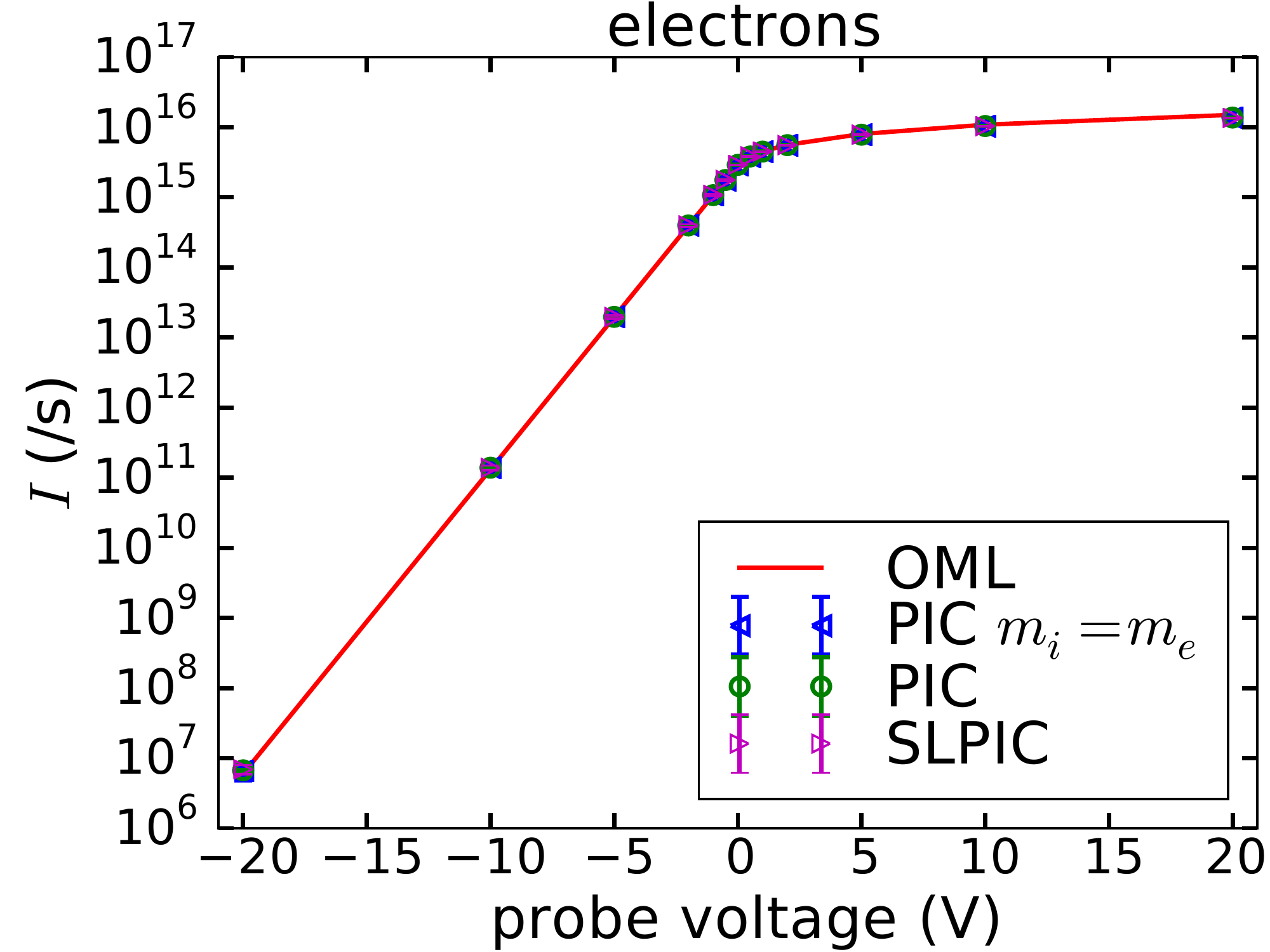}%
\includegraphics*[width=3in]{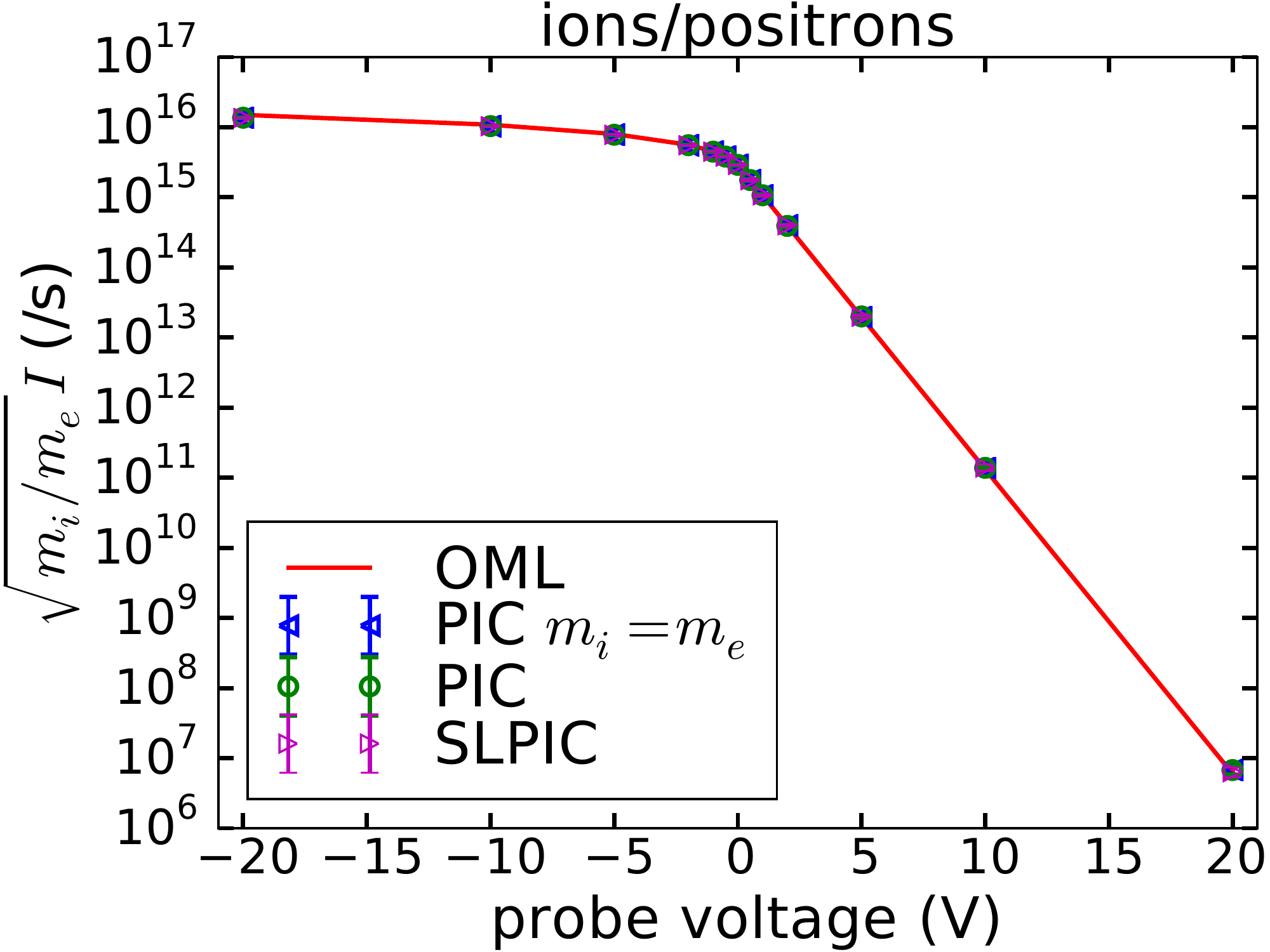}\\
\includegraphics*[width=3in]{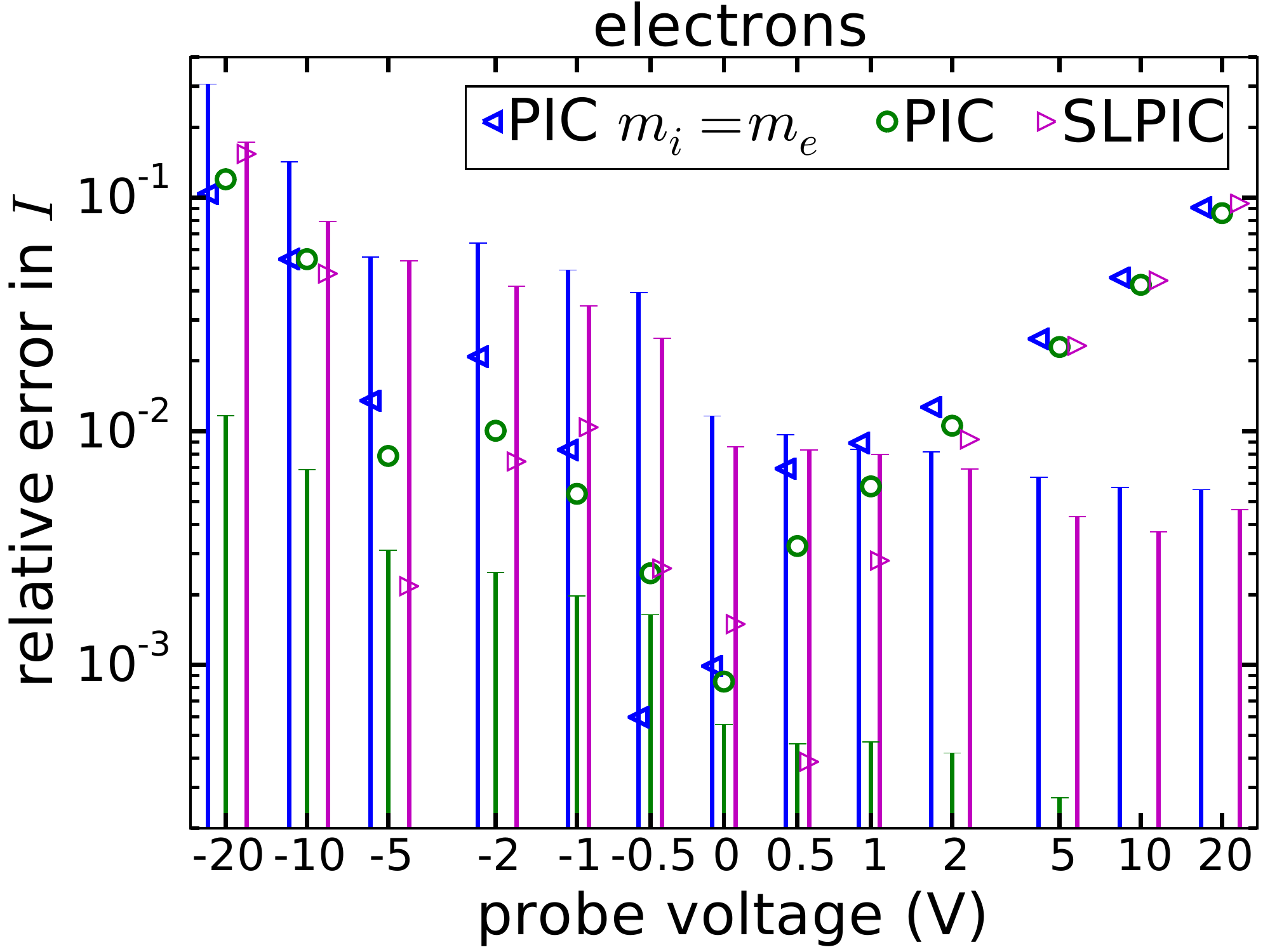}%
\includegraphics*[width=3in]{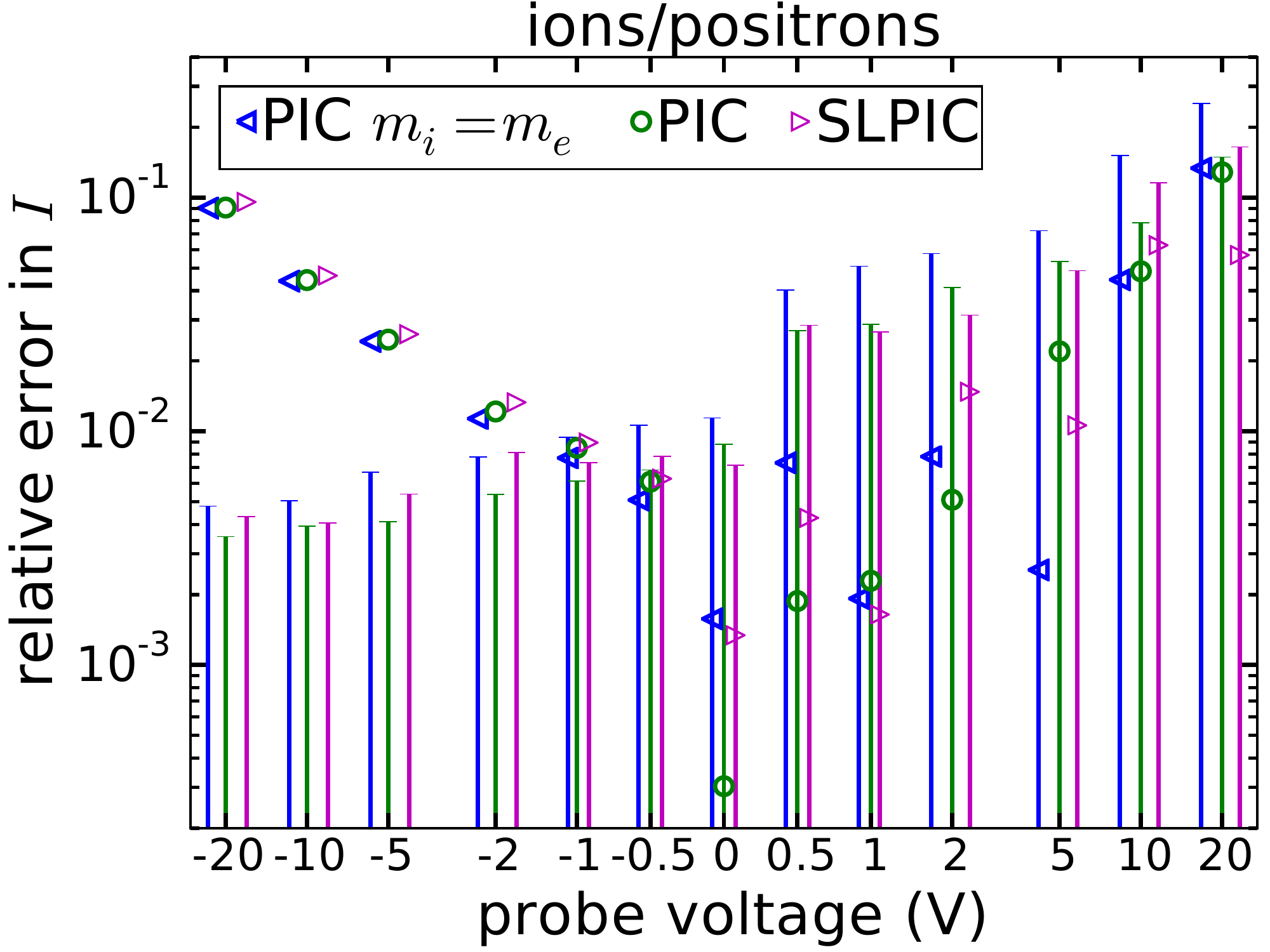}
\caption{\label{fig:results}
Top: The simulated electron current (left) and ion current (right, scaled by
$\sqrt{m_i/m_e}$) hitting the probe, for the three different methods,
and the OML theory \citep[solid line, cf.][]{Allen-1992}.  On these plots, the different
methods are indistinguishable, with error bars too small to see.
Bottom: The relative difference between simulated values and OML theory (symbols),
and vertical bars showing the estimated statistical uncertainty due to shot noise.  Slight offsets in voltage distinguish electron-positron PIC (in blue, offset to the left), electron-ion PIC (in green with no offset), and SLPIC (in magenta, offset to the right).
Symbols above the bars indicate significant differences between simulation and OML theory, due to limited resolution and system size or perhaps to particle trapping; symbols below the bars indicate that the simulation error is less than the statistical uncertainty in measurement of $I$.
In the electron-ion PIC simulation, the statistical uncertainty is much smaller for electrons because many more macro-electrons are simulated over the course of a simulation.
}
\end{figure}

Results are shown in Fig.~\ref{fig:results} for three different methods:
for electron-ion PIC, with the true 
$m_i/m_e\approx 40\times 1836 \approx 7\times 10^4$, for
electron-positron PIC with $m_i/m_e=1$, and for SLPIC with the true
$m_i/m_e\approx 7\times 10^4$.
In each simulation, we count the particles hitting the probe to
calculate the probe current; the positron current must be scaled by
$\sqrt{m_e/m_i}$ to yield the ion current.
The simulations yield essentially identical results; the currents for all three methods are always quite close together compared with the statistical error due to shot noise from random initial positions and velocities of particles.

The only substantive difference, besides speed, is that the electron-ion PIC simulation
naturally simulates many more electrons than ions, and that is reflected in
smaller statistical uncertainties for the electron current (at any one time, the number of macro-electrons and macro-ions in the simulation is the same, but over time, many more macro-electrons are injected and absorbed at the probe).
The statistical uncertainty becomes relatively large when the probe is highly-repelling.  E.g., for $V=-20\:$V, a tiny fraction of the 1~eV macro-electron population can reach the probe; this fraction is increased by our use of variable-weight particles, but even so the number of electrons hitting the probe over the course of the simulation is small.
When the simulation results are above the error bars in Fig.~\ref{fig:results} (bottom), as they are for highly-attracting probe voltages, the difference between simulation and OML theory is statistically significant, and indicates simulation error due to finite $\Delta x$, $\Delta t$, and $L$, or possibly due to particle trapping (cf.~\S\ref{sec:trapping}).
For probe voltages near the plasma temperature, the simulation error is
less than a few percent.  For high repelling voltages, the statistical error is around 10 percent, indicating that the simulation error is not more than that.
For highly-attracting voltages, the simulation error climbs to around 10 percent for $e|V|/T = 20$---less accurate than at lower voltages, but still fairly accurate considering that the measured probe current is many orders of magnitude below the current injected into the simulation.

These methods achieve basically the same accuracy (notwithstanding that PIC yielded lower statistical noise for electrons because it simulated so many of them).
As described above, the standard PIC simulation required $2\times 10^6$ 
timesteps per voltage; depending on the voltage (which affects the number of macroparticles in the simulation) simulations ranged from 390 to 580 core-hours running on 8 cores (of 1 node with two 4-core Westmere processors).
The electron-positron PIC simulations required 1/270 of the number of timesteps,
and ran in 1.6--2.2 core-hours on 8 cores (1 node);
the computation time was thus reduced by a factor of around 250 (nearly 270, as expected).

The SLPIC simulations required only $2/7$ the number of timesteps needed for electron-positron PIC, and ran in 1.4--1.7 core-hours on 8 cores (1 node).
In principle, we believe the SLPIC simulation could run in 2/7 the time required by the pair-PIC simulation, if they were designed to have the same number of macroparticles.
The SLPIC particle-push fundamentally requires more computation than the PIC particle-push; moreover, no attempt has yet been made to optimize the SLPIC push, whereas {\sc VSim}'s normal PIC push has been optimized.
Depending on the machine, it may be possible to optimize the SLPIC push to be competitive with the PIC push (for example, if the particle-push is limited by memory bandwidth and not floating point operations, the extra computation required by SLPIC may add little extra cost).

\section{Particle trapping and convergence to steady-state}
\label{sec:trapping}

The probe current $I(t)$ converged to a steady state in the previously-shown simulations; however, we will see here
that the potential $\Phi(x,y,t)$ did not quite reach a steady state.
Fluctuations in the potential due to stochastic particle noise can reduce particle energy, allowing particles that are attracted to the probe to become trapped in the probe's potential well.
Although the trapping probability is small, the escape probability is similarly small, and thus the population of trapped particles may grow slowly over a long period of time until reaching a steady state \citep{Goree-1992} \citep[see also][]{Zobnin_etal-2000,Lampe_etal-2001,Lampe_etal-2003,Sternovsky_etal-2004}.

In this section, we will see that such a long time is needed that the scaled-mass method is practically essential to study trapping effects.  Here, trapping results from unphysical macroparticle noise, but the scaled-mass method (or SLPIC) would be equally beneficial for studying trapping due to physical Coulomb collisions.  Indeed, even if our PIC code had Coulomb collision capability, we might first want to determine the effect of numerical fluctuations on trapping (as we do here) to determine the minimum number of macroparticles per cell to ensure all trapping effects arise from physical collisions.

An important observation for this section is that PIC simulation cannot reach a true steady state with precisely zero time dependence, but can only achieve a \emph{statistical} steady state with some fluctuations about a constant mean.
These fluctuations can affect the statistical steady state---i.e., the mean value can depend on the fluctuations.
It is to be hoped that, with sufficient resolution and macroparticles, unphysical fluctuations will be small and have a correspondingly small effect on the mean.
However, small fluctuations can potentially have large effects.
For example, small fluctuations can lead to a significant population of particles trapped in a potential well around a Langmuir probe.

In this section, we will show that the potential in our Langmuir probe simulations continues to change---by small amounts, but monotonically---over thousands of ion-crossing times.
We will show that particles attracted to the probe can be trapped in a potential well around the probe for similarly long times, suggesting that the long-term evolution of the potential is caused by the slow accumulation of trapped particles.
Finally, we will briefly investigate the effect of different strength field fluctuations on particle trapping by varying the number of macroparticles per cell.

Particle trapping is most prominent for attracting probe voltages that are large compared with the plasma temperature.
Therefore, we focused on the case with $V=-20\:$V$=-20 T/e$, highly attractive to ions (and highly repelling to electrons).
To study the approach to a steady state,
we ran simulations for very long times, up to $t_f = 8 \times 10^3 L/v_{\rm th,i}$.
To facilitate such long simulations, 
we used $m_i=m_e$ and a coarse resolution: $\Delta x = (1/2)\:$cm, hence just $16^2$ cells.
We ran 3 such simulations with very different average numbers of macro-ions per cell: 7.4, 110, and 1700 macro-ions per cell (the number of macro-electrons was also proportionally increased), expecting cases with more ions/cell to reduce random fluctuations.  We note that other than the resolution, run time, and macroparticle density, these simulations are essentially the same as those described above, including using variable-weight electrons (since they are the repelled species for $V=-20\:V$; see Appendix~\ref{sec:injection}) but constant-weight ions.

\begin{figure}
\centering
\includegraphics*[width=3in]{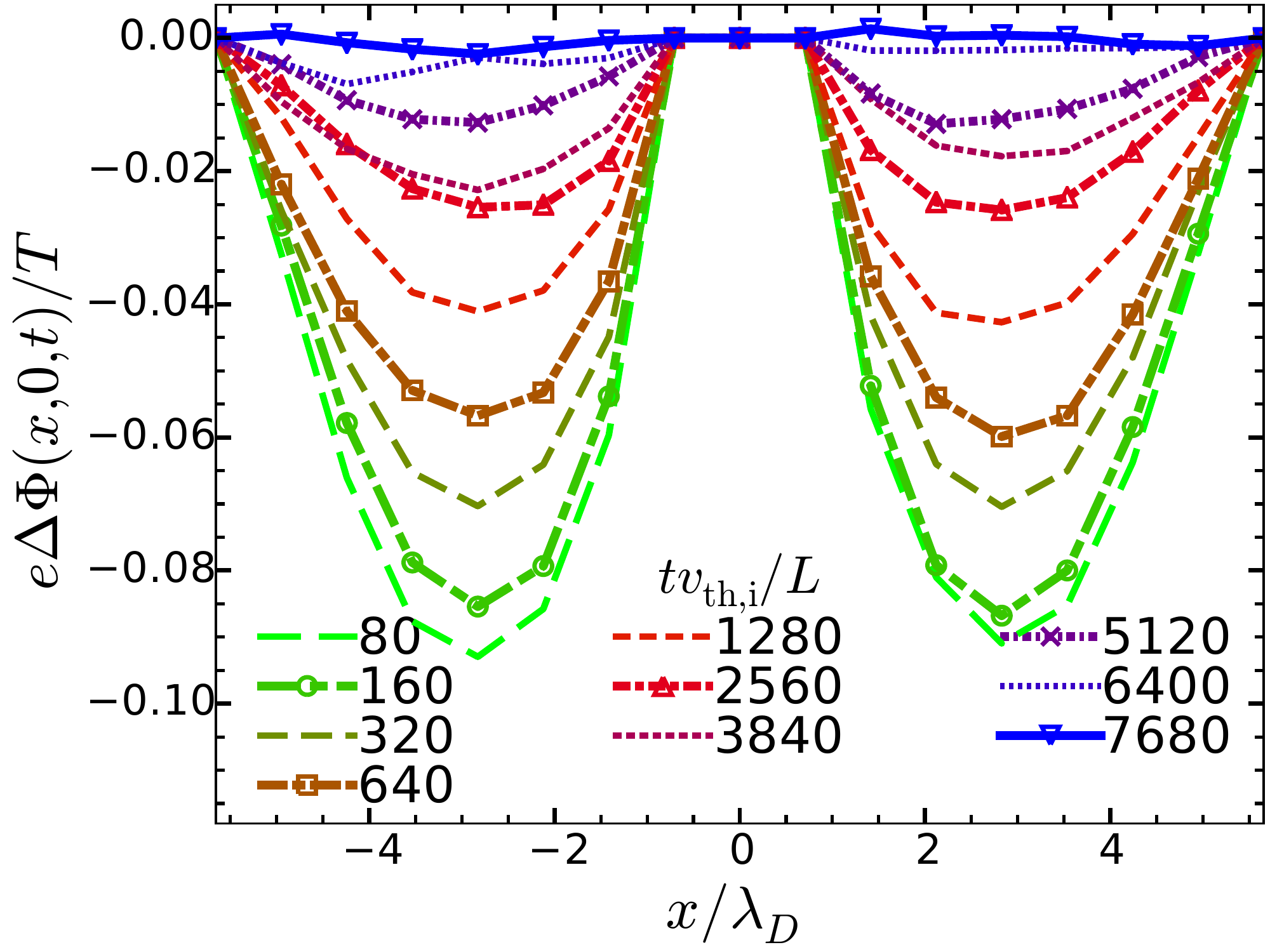}%
\hspace{0.2in}%
\includegraphics*[width=3in]{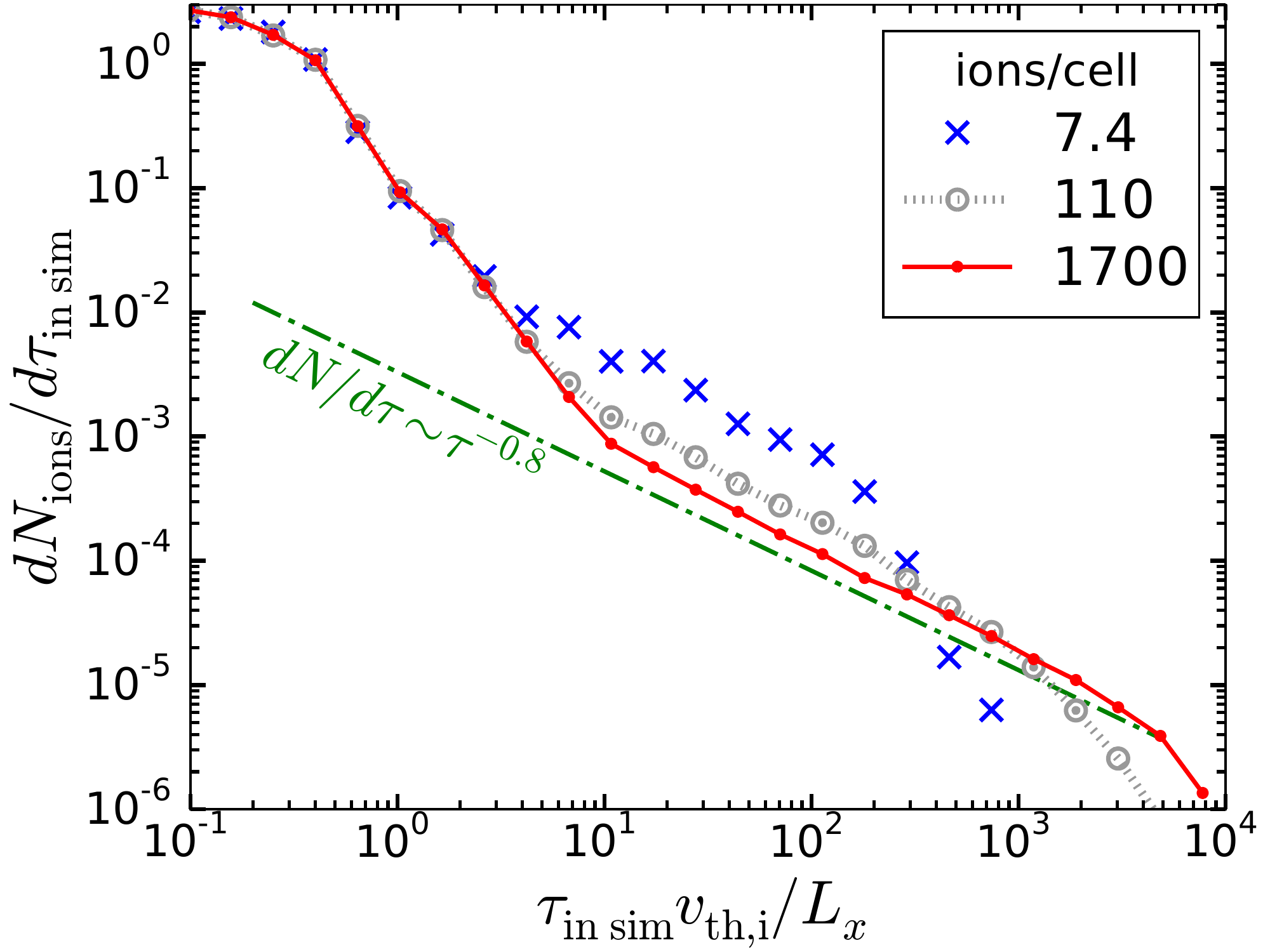}\\
\caption{\label{fig:convergenceAndTrapping}
(Left) The differences $\Delta \Phi = \Phi(x,y=0,t)-\Phi(x,y=0,t_f)$, between the potential at several different times $t$ and the final potential at $t_f=8\times 10^3 \: L/v_{\rm th,i}$, show slow but monotonic change over thousands of ion-crossing times, for a coarse-resolution simulation with around 1700 macro-ions/cell.
The potential is shown along a line through the probe center; the probe boundary condition ensures that the potentials agree exactly for $|x| \leq r_p=0.71\lambda_D$.
(Right) The age-distribution of particles, $dN/d\tau$, for all particles in the simulation at the final time $t_f$, for coarse-resolution simulations averaging 7.4, 110, and 1700 macro-ions per cell.  A particle is ``born'' when injected into the simulation from a boundary at time $t_i$, and its age (at time $t_f$) is $\tau=t_f-t_i$.
Some particles last for thousands of $L/v_{\rm th,i}$ before exiting the simulation or hitting the probe.
}
\end{figure}

After only $3 L/v_{\rm th,i}$, $\Phi(x,y,t)$ is within 0.2~V ($= 0.2 T/e$) of the converged result---good enough for most applications, but $\Phi(x,y,t)$ continues to grow (slightly) for thousands of $L/v_{\rm th,i}$.
Figure~\ref{fig:convergenceAndTrapping} (left) shows the convergence of the potential $\Phi(x,0,t)$, for the lowest-noise simulation with 1700 macro-ions per cell at times $tv_{\rm th,i}/L\in \{$80, 160, 320, 640, 1280, 2560, 3840, 5120, 6400, 7680$\}$, by graphing
$\Delta \Phi(x,0,t) = \Phi(x,0,t) - \Phi(x,0,t_f)$, where $\Phi(x,0,t)$ is a lineout of the potential through the probe center at time $t$, and,
to reduce noise, $\Phi(x,0,t_f)$ is the average of 5 different snapshots at equally-spaced times $t\in [0.96 t_f , t_f]$.
We see that, at late times, $\Phi(x,0,t)$ grows monotonically over time, converging to within $\simeq 0.01\:$V$=0.01T/e$ only after thousands of $L/v_{\rm th,i}$.

The growth of the potential is caused by a slow increase in the number of ions trapped around the attracting probe.
We can see this in Fig.~\ref{fig:convergenceAndTrapping} (right), which
graphs the age distribution of particles ($dN/d\tau$) at time $t_f$ for the three simulations with 7.4, 110, and 1700 macro-ions per cell (on average).
By a particle's ``age'' at time $t_f$, we mean the time $\tau=t_f-t_i$ that a particle has been in the simulation, the time since $t_i$, when the particle first entered the simulation from a boundary ($t_i$ is different for each particle).
Most important, this shows that some ions remain in the simulations for thousands of $L/v_{\rm th,i}$, and therefore one must simulate at least that long to reach a steady state.

Particle trapping cannot happen in a true steady state solution of the Vlasov-Poisson equations for this system.
In a constant potential $\Phi$,
a particle follows a trajectory of constant (kinetic plus potential) energy; either it hits the probe or exits the simulation, in a relatively short time of order $L/v$, where $v$ is the particle's velocity.
However, when $\Phi$ varies in time, a particle may lose enough energy to become trapped in the potential well of the probe.
Indeed, ions with $\tau \gg L/v_{\rm th,i}$ are almost certainly trapped in the potential well of the probe, and so
the steep decline in $dN/d\tau$ around $\tau\sim 1\,L/v_{\rm th,i}$ shows that most ions are not trapped and exit (or hit the probe) in time $\lesssim L/v_{\rm th,i}$.

For $\tau \gtrsim 10 L/v_{\rm th,i}$, $dN/d\tau$ develops into a power law, $dN/d\tau \sim \tau^{-0.8}$, that extends up to some cutoff $\tau_c$.
The power-law of $dN/d\tau$ can be roughly understood with a simplistic model of trapping and escape in which particles execute a random walk in (total) energy due to the fluctuating potential.
That is, in each timestep $\Delta t$, a particle randomly gains or loses energy $\Delta E$ with equal probability.
A particle with total energy $E>0$ can escape the potential well, while a particle with $E<0$ is trapped.
We can then ask: suppose a given particle (soon after injection) walks randomly to $E<0$ at some time $t$; how long will it take the particle to walk randomly back to $E=0$ (at which point it escapes the well)? 
For a simple one-dimensional random walk, the probability that a particle will eventually reach $E=0$ is one; the probability that a particle will reach $E=0$ at time $t+\tau$ (and not before) is proportional to $\tau^{-3/2}$ for $\tau\gg \Delta t$ \citep[cf. the ``first passage time'' for a symmetric 1D random walk, e.g.,][]{Feller-1968}.
Thus the distribution of completed particle lifetimes (mortality age distribution) goes as $\sim \tau^{-3/2}$, 
and hence the distribution of particle ages (for the population of still-living particles at a given time) is $dN/d\tau \sim \tau^{-1/2}$; more precisely, for $\tau \gg \Delta t$, $dN/d\tau \propto (2\pi\tau/\Delta t)^{-1/2}$.

We actually observe $dN/d\tau \sim \tau^{-0.8}$ (Fig.~\ref{fig:convergenceAndTrapping}, right).  The difference between power-law indices
$-1/2$ and $-0.8$ may be caused by the fact that the distribution of random step-sizes $\Delta E$ is not two-valued and constant, as assumed by the model, but continuous and varying in space.  Nearer the probe the electric field is larger and there are fewer particles per cell, both of which should contribute to larger $\Delta E$.
Perhaps more important, the model did not include the possibility of particles hitting the probe.
The additional escape mechanism of hitting the probe is likely to steepen the distribution, consistent with our observations---i.e., particles are less likely to orbit for long times because they might run into the probe.
An orbiting particle may hit the probe simply because the potential does not have exact cylindrical symmetry, and a circular orbit may become increasingly eccentric until it intersects the probe.
However, a particle may also hit the probe because it loses angular momentum due to field fluctuations.

In fact, we see in Fig.~\ref{fig:convergenceAndTrapping} (right) that increasing field fluctuations (by decreasing the number of macroparticles per cell) reduces particle lifetimes.
The cutoff time $\tau_c$ is a rough maximum lifetime, and it appears to increase with the number of macroparticles per cell: 
for 7.4 ions/cell, $\tau_c \sim 200 L/v_{\rm th,i}$; 
for 110 ions/cell, $\tau_c \sim 10^3 L/v_{\rm th,i}$; and for 1700 ions/cell, $\tau_c$ is at least $5\times 10^3 L/v_{\rm th,i}$ and might be limited by the simulation time $t_f$.  A systematic study of the maximum lifetime must be left for future work.

To estimate the importance of accounting for particles hitting the probe,
we measured the fraction of trapped particles that hit the probe (as opposed to escaping through the simulation boundary) as a function of particle lifetime.
For the case of 1700 ions/cell, the fraction of trapped particles that hit the probe (versus escaping) increases from roughly 0.2 to around 0.4 as the lifetime increases from around $10$ to $10^4\:L/v_{\rm th,i}$ (not shown); older particles are more likely than younger particles to end their careers at the probe, but both young and old particles are more likely to escape than to hit the probe.
However, for the simulations with 7.4 and 110 ions per cell, with $\tau_c \ll t_f$, we see that the probability of hitting the probe (versus escaping) rises to 1 as particle lifetime increases past $\tau_c$; i.e., nearly all particles that survive longer than $\sim \tau_c$ end up hitting the probe.
This suggests that finite $\tau_c$ is a consequence of particles hitting the probe; and the increase of $\tau_c$ with the number of ions per cell suggests that field fluctuations are largely responsible for this.

Ion-trapping near Langmuir probes (or, similarly, near negatively-charged dust particles) has been investigated before, although in a different context of trapping due to rare but strong collisions such as charge-exchange collisions where a fast ion collides with a slow neutral particle, resulting in a fast neutral and a slow ion.
A single collision with a neutral can cause an ion to lose enough energy to become trapped in the potential well of the probe; similarly, a collision with a neutral can allow a trapped ion to escape.
\citet{Goree-1992} made the important observation that, because the same collision mechanism is responsible for trapping and untrapping, the number of trapped ions in a steady state could be roughly independent of the collision probability (or mean free path)---and therefore, trapping due to collisions may be important even in the collisionless limit.
However, this applies in the collisionless limit only if there is not 
some other escape mechanism; for example, while a system with perfect cylindrical symmetry might have long-lasting orbits, field fluctuations and asymmetries may cause trapped particles to fall into the probe (e.g., making an orbit increasingly eccentric until it intercepts the probe).
If such effects are much more rapid than collisions (and collisions are very rare), then no large density of trapped ions will build up.

In contrast, this paper considered frequent but small changes to a particle's energy and momentum caused by field fluctuations due to particle noise.
A real plasma, because it contains discrete particles, also experiences field fluctuations, which are essentially responsible for Coulomb collisions;
thus the field fluctuations in our simulations can be viewed as numerical Coulomb collisions (which may or may not mimic real Coulomb collisions).
The effect of frequent, small collisions appears to be at least qualitatively similar to the effect of rare, large-angle collisions \citep{Goree-1992}, with trapping being roughly independent of the collisionality in both cases (again, because trapping and escape are both proportional to the collisionality).
Whereas ion trapping has often been considered distinct from electron trapping, because ions experience different collisions (such as charge-exchange), here the same mechanism applies similarly to electrons as well (when the probe attracts electrons).

At this time, it is worth re-iterating that, although the three different simulation methods proposed here (PIC, scaled-mass PIC, and SLPIC) are all capable of simulating the same steady state, they may have different approaches to that steady state, including different field fluctuations.
This is not really very surprising---even two different PIC simulations differing only in the number of macroparticles per cell can yield different results due to different field fluctuations.
However, we have also seen that these differences are likely to be quite small in most cases.
Although we have shown that, in principle, one might need to use many macroparticles per cell and run for very long times to simulate Langmuir probes precisely, in practice reducing errors to a few percent can be done with $O(10)$ macroparticles per cell and a few $L/v_{\rm th,i}$.
We also point out that to the extent that one is concerned about mimicking the effects of particle trapping in real Langmuir probes, one must determine and consider the most likely trapping mechanisms, including various types of collisions, and which mechanisms are most likely may be rather specific to the particular plasma environment.

\section{Conclusion}
\label{sec:conclusion}

In general, electrostatic PIC simulation of electron-ion plasma can be easily and greatly sped up if a steady-state result is desired.
For such cases, the ion mass $m_i$ can be reduced to $m_i'=m_e$, the electron mass, typically yielding a speed up of a factor $\sqrt{m_i/m_e}$---e.g., 270 times faster for an electron-argon plasma simulation.
This `scaled-mass' simulation can be easily done with most existing PIC codes, without any code changes.
Importantly, the scaled-mass method takes advantage of an exact equivalence (in the steady-state Vlasov-Poisson equations) between two plasma systems with different particle masses---it is not an approximation; the self-consistent steady-state solution from the scaled-mass simulation is as accurate as the traditional electron-ion PIC solution, as long as the ion velocities are properly scaled.

For example, if one wants to start with a drifting Maxwellian ion distribution, with
temperature $T_i$ hence thermal velocity $v_{\rm th,i}=\sqrt{T_i/m_i}$, and drift velocity $v_{di}$, then the scaled-mass simulation must initialize the particles with a Maxwellian with scaled thermal velocity $v_{\rm th,i}' = \sqrt{m_i/m_i'}\, v_{\rm th,i}$ (hence the same temperature~$T_i$), as well as with scaled drift velocity $v_{di}'=\sqrt{m_i/m_i'}\,v_{di}$.
The resulting steady-state (scaled-mass) ion distribution in the simulation
must be rescaled to yield the true ion distribution.

This scaled-mass approach was shown to be equivalent to the numerical timestepping technique of \citet{Serikov_Nanbu-1997,Jolivet_Roussel-2002}, although the scaled-mass approach can be used in a standard PIC code without any code changes.
However, the numerical timestepping method is more straightforward to generalize to phenomena such as collisions, secondary emission, and magnetic fields.
On the other hand, the scaled-mass view is useful because numerical instabilities in the numerical timestepping approach \citep{Jolivet_Roussel-2002} can be understood in terms of known PIC numerical instabilities in an electron-positron plasma.

Numerical timestepping is actually a special case of the more general speed-limited PIC (SLPIC) method \citep{Werner_etal-2018,Jenkins_etal-2021}.  SLPIC also offers straightforward generalization to collisions \citep{Theis_etal-2021}, secondary emission, and magnetic fields (we note that SLPIC has not yet been tested with magnetic fields); additionally, SLPIC offers the potential for greater speed-up, as well as the possibility to simulate time-dependent phenomena.

In this paper, we demonstrated the equivalence of standard PIC, scaled-mass PIC, and SLPIC on simulations of a cylindrical Langmuir probe in an electron-argon plasma, and we took advantage of the faster methods to examine particle trapping around the Langmuir probe and the approach to the steady state.
The scaled-mass and SLPIC methods reduced the computation time by more than two orders of magnitude.
This example demonstrates how these methods could be invaluable in interpreting real Langmuir probe measurements for which analytical methods are inaccurate.  In particular, this work was motivated by the need to understand probes that are not simply planar/cylindrical/spherical or that are not situated within a uniform, homogeneous plasma---such as probes near surfaces in laboratory plasmas or within spacecraft sheaths.  We expect these methods could also be used to speed simulation of probes (or similar problems, such as the charging of dust particles) with complications such as non-Maxwellian or non-uniform plasma, collisions, particle trapping, secondary emission, and possibly even magnetic fields.

\begin{acknowledgments}
This work was supported by the U.S. National Science Foundation grant PHY1707430, 
and by the U.S. Department of Energy SBIR Phase I/II Award DE-SC0015762.
\end{acknowledgments}

\section*{Data Availability}

The data that support the findings of this study are available from the corresponding author upon reasonable request.

\appendix

\section{Particle injection}
\label{sec:injection}

When the probe voltage far exceeds the temperature,
only a tiny fraction of the repelled species (i.e., in the tail of the Maxwellian
distribution) will have sufficient energy to reach the probe.
For example, with an electron temperature of $T=1\:$eV and a 
probe voltage $V = -20\:$V, only a fraction $10^{-8}$ of
electrons have $mv^2/2 \geq -eV_{\rm probe} = 20 T$, or
$v/v_{\rm th} \geq \sqrt{40} \approx 6.3$.
We would therefore need to simulate trillions of electrons to have any hope that a decent sample (thousands) of electrons will hit the probe---assuming, that is, that the simulated macro-electrons have the same velocity distribution as the physical electrons.
In such cases we can use variable-weight particles to over-represent the tail of the distribution, so that a larger fraction of macroparticles can reach the probe.
The ``weight'' of a macroparticle is the number of physical electrons represented by (or contained in) the macroparticle.
To compensate for their over-representation, high-energy macroparticles are given lower weights than low-energy macroparticles.
For example, suppose the velocity distribution of macroparticles is $f_M(v)$, and macroparticles with velocity $v$ have weight $w(v)$; this represents a physical distribution of particles, $f(v)=f_M(v) w(v)$.
If macroparticles with $v_2$ are over-represented compared with $v_1$ [i.e., $f_M(v_2)/f_M(v_1) > f(v_2)/f(v_1)$], then the over-represented macroparticles have lower weights, $w(v_2) < w(v_1)$.
In the following we describe in detail how we injected particles to
obtain sufficient high-energy particles to measure probe currents at
highly-repelling voltages.

The simulations discussed in this paper injected macroparticles from simulation boundaries.
The distribution of injected particles is the
``flux density distribution'' 
$\mathbf{J}(\mathbf{v}) \equiv \mathbf{v} f(\mathbf{v})$;
$\hat{\bf n}\cdot \mathbf{J}(\mathbf{v}) \Delta t$
describes the velocity
distribution of particles crossing the plane with normal
$\hat{\bf n}$ in time $\Delta t$ (per unit area).
Compared with the ``density distribution'' $f(\mathbf{v})$, the
flux distribution is enhanced by $\mathbf{v}$,
because in the given time interval, faster particles travel farther and are
hence more likely to cross the plane.

The flux distribution is given by the physical conditions; we inject particles as if they emerged from a surrounding bath of non-drifting Maxwellian particles with density $n$ and thermal velocity $v_{\rm th}$.
Considering, for concreteness, injection of particles in the $+x$ direction ($v_x>0$) from the $-x$ boundary, the desired flux distribution is
\begin{eqnarray}
  J_x(\mathbf{v}) &=& n v_x (\sqrt{2\pi} v_{\rm th})^{-3}
                    \exp(-v^2/2v_{\rm th}^2) \Theta(v_x)
\end{eqnarray}
where the Heaviside step function $\Theta$ ensures that we inject only
particles with $v_x>0$.

On the other hand,
the flux distribution of injected macroparticles is $J_{Mx}(\mathbf{v})$
[i.e., $J_{Mx}(\mathbf{v})d^3v$ is the number of macroparticles per area per time with velocities in a volume $d^3v$ around $\mathbf{v}$].
Each macroparticle has a weight $w$, which can depend on the macroparticle velocity, $w=w(\mathbf{v})$.
To represent the physical flux correctly, we must have
$w(\mathbf{v}) J_{Mx}(\mathbf{v}) = J_x(\mathbf{v})$.

When injecting constant-weight (cw) macroparticles, all with the same weight $w_{\rm cw}$, then $J_{Mx,\rm cw}(\mathbf{v}) = J_{x}(\mathbf{v})/w_{\rm cw}$.
The physical particle and macroparticle distributions are the same up to a constant, $w_{\rm cw}$.
If we want to inject $M$ particles per timestep $\Delta t$ from the $-x$ boundary of area $L_y L_z$, then
\begin{eqnarray}
  w_{\rm cw} &=& \frac{L_y L_z \Delta t}{M} \int J_x(\mathbf{v}) d^3 v
        \approx \frac{L_y L_z \Delta t}{M} \frac{n v_{\rm th}}{\sqrt{2\pi}}
.\end{eqnarray}
(For a 2D simulation, $L_z=1\:$unit---we use MKS units, so $L_z=1\:$m.)

However, if we want to over-represent particles with $v \gg v_{\rm th}$, 
we can choose $J_{Mx,\rm vw}(\mathbf{v})$ to be independent of $\mathbf{v}$.  
We cannot have a uniform distribution over all $\mathbf{v}$, because it would not be normalizable, so we restrict to a uniform distribution over
$v_x \in [0, v_{\rm max}]$ and $v_y, v_z \in [-v_{\rm max}, v_{\rm max}]$ for some choice of $v_{\rm max} \gg v_{\rm th}$.
We then have to use variable-weight (vw) macroparticles
with velocity-dependent weight $w_{\rm vw}(\mathbf{v})$.
To inject $M$ particles per timestep $\Delta t$ from the $-x$ boundary of area $L_y L_z$, with velocity-independent probability:
\begin{eqnarray}
  J_{Mx,\rm vw}(\mathbf{v}) &=&
              \frac{M}{L_y L_z \Delta t}
   \frac{ 
              \Theta(v_x)
              \prod_{j\in\{x,y,z\}} \Theta(v_{\rm max}-|v_j|)
        }{ 
 \int \Theta(v_x') \prod_{j\in\{x,y,z\}} \Theta(v_{\rm max}-|v_j'|) d^3 v' }
            \nonumber \\
            &=& \frac{ 1}{4v_{\rm max}^3}
              \frac{M}{L_y L_z \Delta t}
              \Theta(v_x)
              \prod_{j\in\{x,y,z\}} \Theta(v_{\rm max}-|v_j|)
              \\
\label{eq:varWeight}
  w_{\rm vw}(\mathbf{v}) &=& \frac{J_x(\mathbf{v})}{J_{Mx}(\mathbf{v})}
    = 
    \frac{n L_y L_z \Delta t}{M} (4v_{\rm max}^3)
    (\sqrt{2\pi} v_{\rm th})^{-3} v_x 
                    \exp(-v^2/2v_{\rm th}^2)
.\end{eqnarray}

While the constant-weight distribution $J_{Mx,\rm cw}$ would not efficiently
simulate currents to a highly-repulsive probe, the variable-weight
distribution $J_{Mx,\rm vw}$ can waste computation on
small-weight particles with negligible contributions.
Therefore, we actually use a mixture of the two distributions,
which we describe next.

Particles are injected as if entering the
simulation from an external Maxwellian ``bath'' of density $n$ at temperature $T$, hence $v_{\rm th}=\sqrt{T/m}$.
During each timestep $\Delta t$, 
we inject (on average) $M$ of particles of charge $q$ from each
side of the simulation.
Again, we describe injection of one species in the $+x$ direction from the
$-x$ side of the simulation, which has area $L_y L_z$;
we repeat the following $M$ times in each timestep $\Delta t$:
\begin{enumerate}
  \item Draw a random velocity $\mathbf{v}_{\rm cw}$ from a 
    Maxwellian flux distribution with thermal velocity $v_{\rm th}$,
    and choose the weight to be 
    $w_{\rm cw} = (nv_{\rm th}/\sqrt{2\pi})L_y L_z \Delta t/M$ 
    (which is the same) for each particle.
    The probability distribution from which velocities are chosen is
\begin{eqnarray}
  p_{\rm cw}(\mathbf{v}) &=& \frac{ v_{x} \exp(-v^2/2v_{th}^2) }{
                          \int v_x \exp(-v^2/2v_{th}^2) d^3 v}
.\end{eqnarray}
  (We note that $v_x$, $v_y$, and $v_z$ can each be generated
  independently, with
  $v_y$ and $v_z$ chosen from a Maxwellian distribution, e.g., by the
  Box-M\"{u}ller transform, and $v_x$ chosen from the above distribution
  proportional to $v_x \exp(-v_x^2/2) v_{\rm th}^2$, by means of, e.g.,
  the inversion method.)

  \item Draw a random velocity $\mathbf{v}_{\rm vw}$ from a 
    uniform distribution $p_{\rm vw}(\mathbf{v})$ 
    within a bounded half-cube 
    $v_x \in [0, v_{\rm max}]$ 
    and $v_y, v_z \in [-v_{\rm max}, v_{\rm max}]$,
    where
\begin{eqnarray}
  v_{\rm max} & \equiv & \textrm{max}\left[
     4 v_{th} , v_{th} + \sqrt{\textrm{max}(0, 2q V_{\rm probe}/m )} 
    \right]
\end{eqnarray}
   This choice of $v_{\rm max}$ ensures that $v_{\rm max} \gg v_{th}$ and $v_{\rm max}$ is greater than the minimum energy needed to reach the probe.
   Then choose a velocity-dependent weight $w_{\rm vw}$
   according to Eq.~(\ref{eq:varWeight}).

  \item Calculate a probability $p_{\rm vw}$ for injecting the variable-weight particle; the probability for injection the constant-weight particle will be $p_{\rm cw}=1-p_{\rm vw}$.
  We choose these probabilities so that high-velocity macroparticles tend to be variable-weight, and low-velocity macroparticles tend to be constant weight.
  For an attracting potential 
    ($qV_{\rm probe} \leq 0$) we do not need variable-weight particles,
    so $p_{\rm vw}\equiv 0$.
    For a repulsive potential ($qV_{\rm probe} > 0$), we always use the variable-weight particle when the particle's energy $W=mv^2/2$ is greater than $2T$ and $W$ is close to energy $qV_{\rm probe}$ needed to reach the probe (i.e., $W \geq qV_{\rm probe}-T$):
\begin{eqnarray}
    p_{\rm vw}(v) & \equiv &
      \textrm{min}\left[ 1, 
       \exp \left(\frac{W - \textrm{max}(q V_{\rm probe} - T, 2T)}{T}
            \right) \right]
            \\
     &=&
     \left\{ \begin{array}{c@{\quad\textrm{for }}l}
       \exp[(W-2T)/T] & W \leq 2T \geq qV_{\rm probe} - T \\
       1 & W \geq 2T \geq qV_{\rm probe} - T \\
       \exp[(W-qV_{\rm probe}+T)/T] & 
                    W \leq qV_{\rm probe}-T \geq 2T \\
       1 & W \geq qV_{\rm probe} - T \geq 2T \\
     \end{array} \right.
     \nonumber
\end{eqnarray}
  As particle energy $W$ increases, $p_{\rm vw}$ increases (up to a maximum of 1).
  \item 
    Inject a particle with velocity and weight 
    $(\mathbf{v}_{\rm vw},w_{\rm vw})$
    with probability $p_{\rm vw}(\mathbf{v}_{\rm cw})$; i.e.,
    draw a uniformly-distributed 
    random number $r \in [0,1]$, and inject the particle
    if $r < p_{\rm vw}(\mathbf{v}_{\rm vw})$ (if not, do nothing).
    Then inject a(nother) particle with velocity and weight 
    $(\mathbf{v}_{\rm cw},w_{\rm cw})$
    with probability $1-p_{\rm vw}(\mathbf{v}_{\rm cw})$,
    again drawing another random number $r' \in [0,1]$
    and if $r' > p_{\rm vw}(\mathbf{v}_{\rm cw})$, inject the particle.
    On average a fraction $p_{\rm vw}(\mathbf{v})$ of the injected
    particles with velocity $\mathbf{v}$ will come from the variable-weight
    distribution, and a fraction $1-p_{\rm vw}(\mathbf{v})$ from 
    the constant-weight distribution.
\end{enumerate}

\setlength{\bibhang}{0.25in}


\end{document}